\documentclass[fleqn,twoside]{article}

\usepackage{axodraw}
\usepackage{color}

\usepackage{espcrc2}
\usepackage{graphicx}

\readRCS
$Id: espcrc2.tex,v 1.2 2004/02/24 11:22:11 spepping Exp $
\usepackage{epsfig}

\newcommand{\AmS}{{\protect\the\textfont2
  A\kern-.1667em\lower.5ex\hbox{M}\kern-.125emS}}

\title{
\vspace*{-18mm}
\rightline{
{\normalsize{LTH 795, DESY 08-092 (July 2008)}}}
\vspace*{+5mm}
Towards the NNLO evolution of polarised parton distributions$\:\!$%
{\thanks{Proceedings of the workshops~ {\it Loops and Legs in Quantum 
Field Theory}, April 2008, Sondershausen (Germany) and (shortened) 
{\it DIS 2008}, London, April 2008.}}
}

\author{A. Vogt\address{Department of Mathematical Sciences, University of 
 Liverpool, Liverpool L69 3BX, United Kingdom},
 S. Moch\address[DESYZ]{Deutsches Elektronensynchrotron DESY, Platanenallee 6, 
 D--15738 Zeuthen, Germany},
 M. Rogal\addressmark[DESYZ]
 and J.A.M. Vermaseren\address{NIKHEF, Kruislaan 409, 1098 SJ Amsterdam, 
 The Netherlands}} 
       
\def\ca{{C_{\!A}}}
\def\cas{{C^{\, 2}_{\!A}}}

\def\cf{{C_F}}
\def\nf{{n^{}_{\! f}}}

\def\cfs{{C^{\:\! 2}_{\! F}}}

\newcommand{\beq}{\begin{equation}}
\newcommand{\eeq}{\end{equation}}
\newcommand{\bea}{\begin{eqnarray}}
\newcommand{\eea}{\end{eqnarray}}
\newcommand{\nn}{\nonumber}
\newcommand{\nin}{\noindent}
\newcommand{\MSb}{$\overline{\mbox{MS}}$}
\newcommand{\as}{\alpha_{\rm s}}
\newcommand{\ass}{\alpha_{\sf s}^{}}
\newcommand{\ars}{a_{\sf s}}

\newcommand{\ra}{\rightarrow}

\newcommand{\ep}{\varepsilon}

\begin{document}

\begin{abstract}
We report on the first calculation of the structure function $g_1^{}$ in 
polarised deep-inelastic scattering to the third order in massless perturbative
QCD.  The calculation follows the dispersive approach already used for the 
corresponding unpolarised cases of $F_{\:\!2,L}$, but additionally involves
higher tensor integrals and the Dirac matrix $\gamma_5^{}$ in $D \neq 4$ 
dimensions.  
Our results confirm all known two-loop expressions including the coefficient 
functions of Zijlstra and van Neerven not independently verified before.  
At three loops we extract the helicity-difference next-to-next-to-leading order 
(NNLO) quark-quark and gluon-quark splitting functions $\Delta P^{}_{\rm qq}$ 
and $\Delta P^{}_{\rm qg}$. The results exhibit interesting features concerning
sum rules and the momentum-fraction limits $x \ra 1$ and $x \ra 0$.
\vspace*{-1mm}
\end{abstract}

\maketitle

\section{Introduction}
 
Two decades after the seminal EMC measurement \cite{Ashman:1987hv} set off the
so-called spin crisis, the internal spin structure of the nucleon remains a
very active field of research. By now the measurements of polarised (spin-%
dependent) deep-inelastic scattering (DIS) -- the very process first measured
by 

\begin{figure}[htbp]
\vspace*{-0.9cm}
\centerline{\hspace*{2mm}\epsfig{file=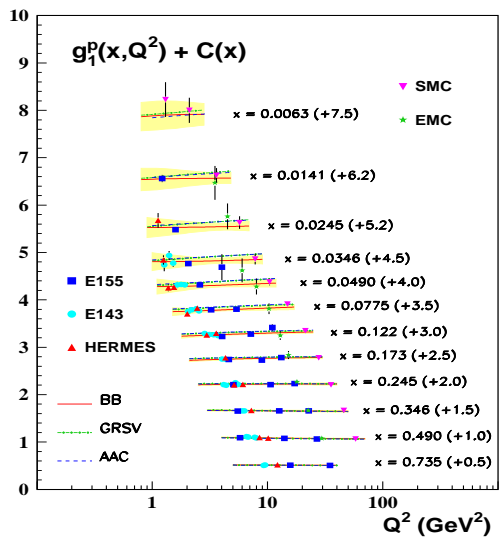,width=6.5cm,height=7.0cm}}
\vspace{-9mm}
\caption{\label{g1p-data}
The end-2006 world data on the proton structure function $g_1^{}$ with some NLO QCD 
fits~\cite{g1p-plot}} 
\vspace*{-1mm}
\end{figure}

\nin
EMC -- have reached a high accuracy but are still restricted to relatively
small scales $Q^2$, see Fig.~1.

\vspace{1mm}
Dramatic further improvements, in particular of the kinematic coverage, can be 
expected from the planned Electron-Ion Collider, see Ref.~\cite{EIC-WP}.
But, as rather low values of $Q^2$ imply rather large QCD corrections, even 
analyses of available data will profit from calculations beyond the present 
next-to-leading order (NLO) approximation.  

\section{Polarised DIS in perturbative QCD} 

The kinematics and perturbative-QCD factorization of inclusive DIS are 
recalled in Fig.~2.

\begin{figure}[htbp]
\vspace*{-0.55cm}
\parbox{3.7cm}{
\begin{center}
\vspace*{-1.4cm}
\includegraphics[bb = 170 560 430 360, scale = 0.4]{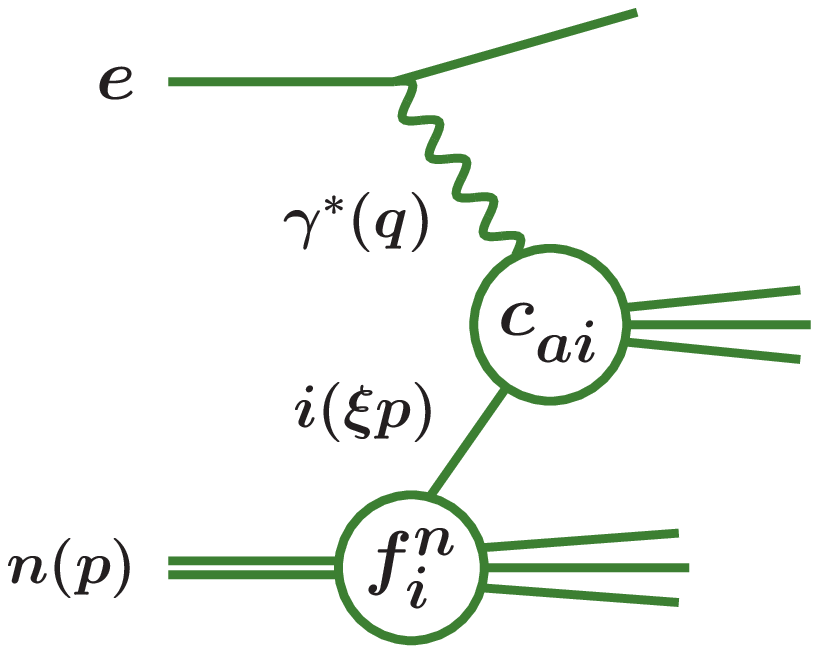}
\end{center}
}
\parbox{3.5cm}{
\vspace{1mm}
{Scale, Bjorken variable}
\begin{eqnarray*}
 ~~~Q^2\!\!  & = &  - q^2 \\[0.5mm]
 ~~~x        & = &  Q^2/(2\:\! p\cdot\! q)
\end{eqnarray*}
\vspace{1mm}
Lowest order$\,$: {$x = \xi$}}
\vspace*{-5.5mm}
\caption{\label{DISkin}
DIS in the QCD-improved parton~model$\!$}
\vspace*{-6mm}
\end{figure}

The measurements of the difference
$\:\sigma_{\stackrel{\scriptscriptstyle\rightarrow}{e\,}
           \stackrel{\scriptscriptstyle\rightarrow}{n} }
 - \sigma_{\stackrel{\scriptscriptstyle\rightarrow}{e\,}
           \stackrel{\scriptscriptstyle\leftarrow}{n} }\:$
of the helicity-dependent cross sections yield the polarised structure function
$\,g_1^{\:\!n}$ of the nucleon~$n$. \linebreak
In perturbative QCD, neglecting contributions suppressed by $1/Q^2$, this 
quantity is given by
\bea
  g_1^{\:\!n}(x,Q^2) \!\! &\!\!=\!\!& \!\!\!\!\int_x^1 \! \frac{d\xi}{\xi}
  \, c_{g_1^{}\!,k}\Big( {x \over \xi}, \ass(\mu^2),
  {\mu^2 \over Q^2} \Big) \Delta f_{k}^n(\xi,\mu^2) \!\!
  \nn \\[-2mm] & & 
\label{eq:g1n}
\eea
\nin
where, as in Eq.~(\ref{eq:evol}) below, the sum over the parton species $k$ is
understood. $c_{g_1^{}\!,k}$ represents the corresponding coefficient
functions (mass-factor\-ized partonic cross sections) for $g_1^{}$, $\ass\!$ 
denotes the strong coupling, and $\mu \!=\! {\cal O}(Q)$ stands for the 
renormalization and mass-factorization scale. The helicity-difference parton 
distributions $\,\Delta f_{i\,} = f_{i\ra} - f_{i\leftarrow\,}$ obey
the evolution equations
\beq
  \frac{d}{d \ln \mu^2} \, \Delta f_i^{}(\mu^2)
   = \left[ {\Delta P^{}_{ik}(\as(\mu^2))}
  \otimes \Delta f_k^{}(\mu^2) \right]
\label{eq:evol}
\eeq
where $\otimes$ denotes the Mellin convolution written out in Eq.~(\ref
{eq:g1n}). The initial conditions for Eq.~(\ref{eq:evol}) are, of course,
incalculable in perturbative QCD. Present lattice techniques can provide, at 
best, only a few low Mellin moments. Hence predictions for, e.g., processes 
at RHIC, can be obtained only via fit-analyses of reference observables such 
as $g_1^{}$ and the universality of the parton densities. For the present
state-of-the-art see, e.g., Ref.~\cite{deFlorian:2008mr}.

\vspace{1mm}
The perturbative quantities entering a general hard-scattering observable $a$ 
are thus the splitting functions $P$ and the coefficient functions $c^{}_{a}$,
\begin{eqnarray}
 ~P \, &\!\! =\!\! &
    \ass\, P^{(0)} \, +\, \as^2\, P^{(1)}
    \, +\, \as^3\, P^{(2)} +\, \ldots \nn \\[0.5mm]
 ~c^{}_{a} &\!\! =\!\! & \underbrace{ \as^{\,n_a}
   \Big[ \, c_a^{(0)} + \as\, c_a^{(1)} }_{\rm NLO} \:\!
    + \:\, \as^2\, c_a^{(2)} \,\, +\, \ldots \,\Big] \: .
\label{eq:Pexp}
\end{eqnarray}
Being the first serious approximation, the NLO cannot provide a sound estimate 
of the error from the truncation of the perturbation series. The NNLO requires
the two-loop coefficient functions for all observables under consideration and
the universal three-loop splitting functions. The latter quantities 
$\Delta P^{(2)}_{ik}$ for the polarised evolution (\ref{eq:evol}) are the 
main subject of the present research.

\section{Previous second-order calculations}

Before we turn to our new third-order calculations, it is useful to briefly 
recall the previous two-loop results for the coefficient functions for
$g_1^{}$ and the splitting functions $\Delta P^{}_{ik}$. All those calculations
have been carried out in dimensional regularization. Hence the helicity-%
difference projection of initial quarks introduces the issue of $\gamma_5^{}$
in $D \neq 4$ dimensions into the problem.   

\vspace{1mm}
The perturbative corrections for the polarised structure function $g_1^{}$ to 
order $\as^2$ have been calculated 15 years ago by Zijlstra and van 
Neerven~\cite{Zijlstra:1993sh}. This $x$-space calculation, carried out using 
the so-called Larin scheme \cite{g5-Larin} for $\gamma_5^{}$ (see below), 
yielded the NNLO coefficient functions $c_{g_1,\,\rm q/g}^{(2)}$ and well as 
two of the four polarised NLO splitting functions: $\Delta P_{\rm qq}^{(1)}$ 
and $\Delta P_{\rm qg}^{(1)}$.

\vspace{1mm}
The complete set of NLO splitting functions $\Delta P_{ij}^{(1)}$ was first 
computed two years later by Mertig and van Neerven using the operator-product 
expansion \cite{Mertig:1995ny}. Here $\gamma_5^{}$ was treated using the 
reading-point method of Ref.~\cite{Korner:1991sx}. 
The results were independently verified already a few months later by 
Vogelsang \cite{Vogelsang:1995xx} in an axial-gauge calculation. This 
calculation was performed using the original HVBM prescription for 
$\gamma_5^{}$ \cite{g5-HVBM} but included checks with the other schemes
mentioned above.

\vspace{1mm}
All three theoretically (but not computationally) identical prescriptions for
$\gamma_5^{}$ require an additional renormalization or factorization of the 
loop results in Refs.~\cite{Zijlstra:1993sh,Mertig:1995ny,Vogelsang:1995xx}. 
The complete set of the corresponding expressions for $g_1^{}$ to NNLO has been 
derived in Ref.~\cite{Matiounine:1998re} ten years ago. 
We will return to this issue below.

\section{The three-loop computation via forward Compton amplitudes}

The present calculation follows the approach used before in our computation of 
the unpolarised splitting functions \cite{MVV-Pij} (see also Ref.~\cite
{MVV-proc}) and the coefficient functions for $F_{2,L}$ \cite{MVV-F2L} to order
$\as^3$. Thus we employ the optical theorem

\hspace*{-6mm}\includegraphics[bb = 70 425 100 535, scale = 0.5]{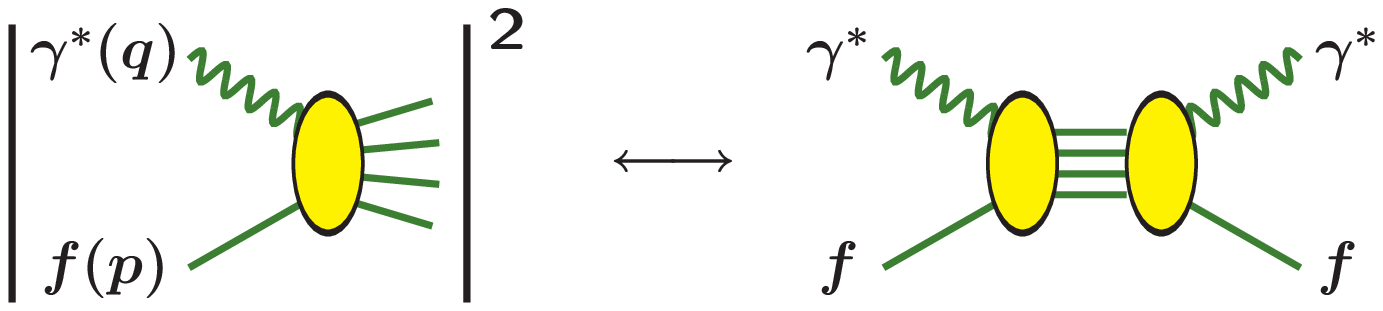}

\nin
to transform the $\gamma^\ast\! f$ total cross sections into forward Compton 
amplitudes. Due to a dispersion relation in the partonic Bjorken variable $x$,
the coefficient of $(2\:\!p\!\cdot\! q)^N$ provides the $N$-th moment
$$
  ~~\hat{F}_a^N \: = \: {\textstyle \int}_{\!0}^1 \, dx \: x^{N-1} \hat{F}_a(x)
$$
of the partonic structure function $\hat{F}_a$ selected by a suitable 
projection of the $\gamma^\ast$-amputated graphs.
 
In particular, the projection of the partonic tensor $\widehat{W}^{\mu\nu}$ on 
$\hat{g}_1^{}$ in $D$ dimensions reads~\cite{Zijlstra:1993sh}
\beq
  ~~\hat{g}_1^{} \,=\, 2\, [(D-2)(D-3) (p\!\cdot\! q)]^{-1}\,
  \ep_{\mu\nu pq}\, \widehat{W}^{\mu\nu} 
\label{eq:g1proj}
\eeq
using the {\sc Schoonschip} notation for contracted indices.
The N$^{\,l-1}$LO splitting functions $\Delta P_{\rm qf}^{(l-1)}$ \\[-1mm]
and the N$^{\,l}$LO coefficient functions $c_{g_{1\!}^{},\rm \:\!f}^{\,(l)}$ 
can be 
\linebreak\\[-4mm]
extracted from the $1/\ep$ poles ($D = 4 - 2 \ep$) and $\ep^0$ parts, 
respectively, of the corresponding $l$-loop results for 
$\gamma^\ast\!f\ra\gamma^\ast\!f$, cf.~Refs~\cite{MVV-Pij,MVV-proc,MVV-F2L}. 

\subsection*{Treatment of the integrals}

\vspace{1mm}
In order to illustrate the computation of integrals required to evaluate the 
forward Compton amplitudes, let us consider one of the hundreds of
$\gamma^\ast\! g$ three-loop diagrams shown here together with a pictorial 
representation of its momentum~flow:

\vspace*{-1.5cm}
\begin{center}\hspace*{-6mm}
\SetScale{0.55}
\begin{picture}(300,90)(-30,-25)

\SetColor{Red}  

\SetWidth{3}
\Line(35,0)(0,0)
\Line(70,0)(35,0)
\Line(105,0)(70,0)
\SetWidth{2}
\Gluon(-15,-15)(0,0){3}{2}
\Gluon(105,0)(120,-15){3}{2}

\SetColor{OliveGreen} 
\Gluon(70,0)(70,60){4}{5} 
\Gluon(35,0)(35,60){4}{5} 

\Photon(105,60)(120,70){2}{2}
\Photon(0,60)(-15,70){2}{2}

\Vertex(0,0){2}      
\Vertex(0,60){2}      
\Vertex(105,0){2} 
\Vertex(35,0){2}
\Vertex(70,0){2}
\Vertex(32,60){2}  
\Vertex(70,60){2}  
\Vertex(105,60){2}

\Line(0,0)(0,60)
\Line(0,60)(35,60)
\Line(35,60)(70,60)
\Line(70,60)(105,60)
\Line(105,60)(105,0)

       \Line(250,55)(300,55) \Line(250,5)(300,5)
       \Line(250,55)(250,5) \Line(300,55)(300,5)
       \CArc(250,30)(25,90,270) \CArc(300,30)(25,270,90)
       \Line(225,30)(215,30) \Line(325,30)(335,30)
{\SetColor{Red} \SetWidth{5} \Line(300,55)(250,55) 
 \CArc(250,30)(25,90,135)\CArc(300,30)(25,45,90) 
 }

\SetColor{Black}
\SetScale{0.8}
\SetWidth{0.8}
\LongArrow(105,20)(125,20)

\end{picture}
\end{center}
\vspace{-8mm}

\nin
For the latter one disregards the external parton lines and draws the remaining
self-energy type diagram, the topology of which is denoted following the 
notation of Refs.~\cite{MINCER}. Our example is a ladder (LA) diagram.
The (partly additional) denominators carrying the parton momentum $p$ are then 
indicated by the fat (in the coloured version: red) lines. Here $p$ runs, after
turning the diagram upside-down, through the lines 1, 2 and 3. Thus the example
is assigned the subtopology LA$_{13}$. 

\vspace{1mm}
According to our above discussion, we need analytic expressions for the 
(dimensionless) coefficients $I(N)$ of $(2\:\!p\cdot\! q)^N / Q^{2\alpha}$. One
might try to obtain $I(N)$ by brute force, Taylor-expanding the denominators 
with $p$ and working out the sums. It turned out that such a strategy, in 
general, does not work. Instead, we employed identities based on integration by
parts, scaling arguments and form-factor decompositions (see Sect.~2 of 
Ref.~\cite{MVV-Pij}$_1$) to successively simplify the integrals.

\vspace{1mm}
The LA$_{13}$ integrals, e.g., can be simplified by applying $\,p^{\:\!\mu}
\partial / \partial q^{\:\!\mu}$ both inside and outside the integral. For the 
scalar integral with unit denominators this yields \pagebreak
 
\vspace*{1mm}
\raisebox{-25pt}{  
\SetScale{0.4} 
\hspace*{-9mm}\begin{picture}(50,20)(-10,-5)
\SetColor{OliveGreen} 
\SetScale{0.5}
\SetWidth{2}
       \Line(35,70)(85,70) \Line(35,20)(85,20) 
       \Line(35,70)(35,20) \Line(85,70)(85,20) 
        \CArc(35,45)(25,90,270) \CArc(85,45)(25,270,90)
        \Line(10,45)(0,45) \Line(110,45)(120,45)
        {\SetColor{Red} \SetWidth{4} 
        \CArc(35,45)(25,90,135) \CArc(85,45)(25,45,90)
        \Line(35,70)(85,70)} 
\SetColor{Black}
        \PText(60,81)(0)[l]{1} 
        \PText(60,13)(0)[l]{1} 
        \PText(40,45)(0)[l]{1} \PText(80,45)(0)[r]{1}
        \PText(5,60)(0)[l]{1}\PText(20,72)(0)[lb]{1}
        \PText(115,60)(0)[r]{1}\PText(100,72)(0)[rb]{1}
        \PText(15,20)(0)[l]{1} \PText(105,20)(0)[r]{1}
\Text(70,22)[]{+} 
\end{picture} 
}

\vspace*{-18mm}
$$ \hspace*{27mm}
 \frac{N\!+\!3\!+\!3\epsilon}{N\!+\!2}\, \frac{2p\!\cdot\! q}{q^2}
\hspace*{-1mm}
\raisebox{-23pt}{  
\SetScale{0.4}
\SetColor{OliveGreen} 
\begin{picture}(50,20)(0,-3)
\SetScale{0.5}
\SetWidth{2}
       \Line(35,70)(85,70) \Line(35,20)(85,20) 
       \Line(35,70)(35,20) \Line(85,70)(85,20) 
        \CArc(35,45)(25,90,270) \CArc(85,45)(25,270,90)
        \Line(10,45)(0,45) \Line(110,45)(120,45)
        {\SetColor{Red} \SetWidth{4} 
        \CArc(35,45)(25,90,135) \CArc(85,45)(25,45,90)
          \Line(35,70)(85,70)} 
\SetColor{Black}
        \PText(60,81)(0)[l]{1} 
        \PText(60,13)(0)[l]{1} 
        \PText(40,45)(0)[l]{1} \PText(80,45)(0)[r]{1}
        \PText(5,60)(0)[l]{1}\PText(20,72)(0)[lb]{1}
        \PText(115,60)(0)[r]{1}\PText(100,72)(0)[rb]{1}
        \PText(15,20)(0)[l]{1} \PText(105,20)(0)[r]{1}
\end{picture} 
}
$$

\vspace{-3mm}
$$
\hspace*{9mm}\!\!
=  \frac{2}{N\!+\!2}
\raisebox{-23pt}{  
\SetScale{0.4} 
\begin{picture}(50,20)(0,-3)
\SetScale{0.5}
\SetWidth{2}
\SetColor{OliveGreen} 
       \Line(35,70)(85,70) \Line(35,20)(85,20) 
       \Line(35,70)(35,20) \Line(85,70)(85,20) 
        \CArc(35,45)(25,90,270) \CArc(85,45)(25,270,90)
        \Line(10,45)(0,45) \Line(110,45)(120,45)
        {\SetColor{Red} \SetWidth{4} 
        \CArc(35,45)(25,90,135) 
          \Line(35,70)(85,70)} 
\SetColor{Black}
        \PText(60,81)(0)[l]{1} 
        \PText(60,13)(0)[l]{1} 
        \PText(40,45)(0)[l]{1} \PText(80,45)(0)[r]{1}
        \PText(5,60)(0)[l]{1}\PText(20,72)(0)[lb]{1}
        \PText(105,70)(0)[rb]{2}
        \PText(15,20)(0)[l]{1} \PText(105,20)(0)[r]{1}
\Text(95,22)[]{(5)} 
        \end{picture} 
}
$$
\addtocounter{equation}{1}

\vspace*{-3mm}
\nin
Here the LA$_{13}$ integral occurs twice, once with a prefactor $2\:\!p\cdot\! 
q$. Hence Eq.~(5) represents a difference equation (here of order $n=1$)  
expressing its coefficient $I(N)$ in terms of that of a LA$_{12}$ integral with 
an enhanced denominator in the~3-line,
\beq
\label{eq:diff}
 ~~a_0(N) I(N) - \dots  - a_n(N) I(N\! -\! n) \,=\, G(N) \: . \;
\eeq
First-order recursion relations like Eq.~(5) can be reduced to a sum.
Higher-order recursions (we needed equations up to $n=4$) can be solved by
inserting a suitable ansatz into Eq.~(\ref{eq:diff}). Both procedures exploit 
the fact that all integrals required for the computation of the splitting
functions can be expressed in terms of harmonic sums~\cite{Vermaseren:1998uu}.

\vspace{1mm}
Despite being uncharacteristically simple in both derivation and size, Eq.~(5) 
illustrates the strict hierarchy of subtopologies in our procedure. Our 
LA$_{13}$ example can be evaluated only once the LA$_{12}$ integral in Eq.~(5) 
is known. This integral requires the so-called basic building blocks 
(with only one $p$-dependent denominator) LA$_{11}$ and LA$_{22}$ together with
other integrals of simpler topologies where a non-$p$ denominator has been 
removed. Also those integrals need to be evaluated in terms of yet simpler
cases, and so on.
 
\vspace{1mm}
During the first half of this decade, reduction chains for all subtopologies 
were constructed and coded in {\sc Form} \cite{FORM}. This and the computation 
of all integrals (and their much more numerous sub-integrals) entering the 
diagrams for the unpolarized cases \cite{MVV-Pij,MVV-proc,MVV-F2L} took 
years of both human and computing resources. 
It would not have been possible to get through without extensive tabulation of 
intermediate results. By 2005, a database had been accumulated of more than 
100$\,$000 integrals requiring about 3.5 GBytes of disk~space.

\vspace{1mm}
However, as we will see below, this database is still not large enough to cover
all integrals required for the three-loop calculation of the spin-dependent 
structure function $g_1^{}$.

\subsection*{Numerators for a non-planar diagram}

\vspace{1mm}
One of the more difficult three-loop diagrams for the $\gamma^\ast q$ Compton 
amplitudes is shown below in fully amputated form:

\vspace{-0.5mm}
\raisebox{-60pt}{\hspace*{0.5mm}
\begin{picture}(50,20)(0,-5)
\SetColor{OliveGreen}
\SetScale{0.7}
\SetWidth{1.6}
       \Gluon(60,70)(85,70){3}{2.5}
       \Gluon(35,20)(85,20){-3}{5.5}
       \Line(53,52)(85,20)
       \Line(85,70)(63,48)
       \Line(57,42)(35,20)
       \CArc(35,45)(25,90,270) \CArc(85,45)(25,270,90)
       \CArc(35,45)(25,90,135) \CArc(85,45)(25,45,90)
       {\SetColor{Red} \SetWidth{2}
       \Gluon(35,70)(53,52){-3}{2.5}
       \Line(35,70)(60,70)}
\SetColor{Black}
       \Vertex(60,70){2} \Vertex (53,52){2}
       \Vertex(10,45){2.5} \Vertex(110,45){2.5}
       \Text(0,31)[c]{$\nu$}
       \Text(95,30)[c]{$\mu \quad \sim$}
       \Text(194,15)[c]{$(7)$}
        \PText(60,9)(0)[l]{5} \PText(60,82)(0)[l]{2}
        \PText(35,54)(0)[l]{7} \PText(80,54)(0)[r]{8}
        \PText(13,74)(0)[l]{1} \PText(106,74)(0)[r]{3}
        \PText(13,19)(0)[l]{6} \PText(106,19)(0)[r]{4}
\end{picture}
}

\vspace*{-5mm}
$$
 \!\frac {\gamma(i_1^{},p_7^{},i_2^{},p_4^{},\mu,p_3^{},i_3^{},p_8^{},i_2^{},
 p_6^{},\nu,p_1^{},i_1^{},p\!+\!p_2^{},i_3^{})}
 {p_1^{\:\!2} \ldots p_8^{\:\!2}\, (p+p_2^{})^2 (p+p_7^{})^2} 
$$
\addtocounter{equation}{1}
Here the numerator has been written in the notation used by {\sc Form}~\cite
{FORM}, and the numbers in the graph correspond to the line numbering 
convention in {\sc Mincer}~\cite{MINCER}.
Thus this diagram, with the quark entering in line 2 and leaving in 
line 7, is of the subtopology NO$_{27}$. 

\vspace{0.5mm}
In order to determine the unpolarised splitting function, it is sufficient
to evaluate (7) $\!\cdot\: \gamma(p)\,g_{\mu\nu\,}^{}$. The resulting trace
leads to numerators
$$
(p_2^{}\!\cdot\! p)^{k_2} (p_3^{}\!\cdot\! p)^{k_3}
(p_2^{}\!\cdot\! q)^{k_9} \mbox{~~with~~} k_2 + k_3 + k_9 \leq 3 
$$
and the denominator $1/p_2^{\:\!2}$ cancels in all terms, which already 
reduces the number of required reduction steps by one.
 
\vspace{0.5mm}
The disentangling of the contributions to the coefficient functions for $F_2$ 
and $F_L$ required the additional calculation of (7) $\!\cdot\: \gamma(p)\, 
p_\mu p_\nu / (p\!\cdot\! q)^2$ which includes higher numerators, 
$$
(p_2^{}\!\cdot\! p)^{k_2} (p_3^{}\!\cdot\! p)^{k_3}
(p_2^{}\!\cdot\! q)^{k_9} \mbox{~~with~~} k_2 + k_3 + k_9 \leq 4 \, .
$$
Again no denominator $1/p_2^{\:\!2}$ remains in any term.

\vspace{0.5mm}
The determination of the spin splitting function and the coefficient function
for $g_1^{}$ requires the evaluation of (7) $\cdot\: \gamma(p,5)\,
\ep_{\mu\nu pq}/(p\cdot\! q)\,$ with the $\gamma$-term arising from the
quark helicity-difference and the $\ep$-tensor from the $g_1^{}$ projection 
in Eq.~(\ref{eq:g1proj}). Now one encounters yet higher numerators
$$
(p_2^{}\!\cdot\! p)^{k_2} (p_3^{}\!\cdot\! p)^{k_3}
(p_2^{}\!\cdot\! q)^{k_9} \mbox{~~with~~} k_2 + k_3 + k_9 \leq 5 
$$
and the denominators $1/p_2^{\:\!2}$ remain in terms up to the highest 
numerators.  Thus, compared to the previous calculations of $F_2$ and $F_L$ 
\cite{MVV-Pij,MVV-proc,MVV-F2L}, two additional reduction steps (often leading 
to additional denominator enhancements, recall Eq.~(5)) are required in the 
polarised case.

The situation is similar in other difficult sub\-topologies including 
NO$_{12\,}$, BE$_{68\,}$ and LA$_{17}$ (some of which, unfortunately, are far
from being fully automated), leading to a large number of new (sub-)integrals
beyond the 2005 database.

\vspace{1mm}
Using codes and experience from the previous unpolarized calculations, we were
able to compute the required integrals within a couple of months. This would 
have been impossible without the possibility of fixed-$N$ checks at all 
intermediate stages facilitated by the {\sc Mincer} program~\cite{MINCER}.

\subsection*{Treatment of {\boldmath $\gamma_{5}^{}$} and two-loop checks}

\vspace{1mm}
We employ the Larin prescription for the quark helicity-difference projector, 
i.e., we use \cite{g5-Larin}
\beq
  ~~~p\!\!\!\!\: / \:\! \gamma_{5,L}^{} \:=\: {i \over 6}\:
  \ep_{p\mu\nu\rho}\, \gamma_\mu \gamma_\nu \gamma_\rho \:\: .
\eeq
Hence, after coupling-constant renormalization and mass-factorization, we need
to perform a scheme transformation $Z$ which nullifies the first moment of the
non-singlet (ns) quark-quark splitting function in accordance with the conservation 
of the non-singlet axial vector current \cite{Mertig:1995ny,Vogelsang:1995xx},
\bea
  ~~~g_1^{} &\!\! =\!\! & c_{\:\!g_1^{},L} \,\Delta f_L^{} \nn\\
            &\!\! =\!\! & c_{\:\!g_1^{},L} \, Z^{-1} Z \,\Delta f_L^{}
          \: = \: c_{\:\!g_{1\!}^{},\scriptscriptstyle\overline{\rm MS}}^{}
                      \:\Delta f_{\,\scriptscriptstyle\overline{\rm MS}}^{}
  \; .
\label{eq:g1trf}
\eea
The general form of the corresponding transformation of the splitting functions
to NNLO in the flavour-singlet case, where Eq.~(\ref{eq:g1trf}) has to be read 
as a matrix equation, can be found, e.g., in Sect.~2 of Ref.~\cite
{vanNeerven:2000uj}. For the special case
\beq
 ~Z_{ij} = \delta_{ij} + \ars z_{qq}^{(1)} \delta_{qq} + \ars^2 
 \big( z_{qq,\rm ns}^{(2)} + z_{qq,\rm ps}^{(2)} \big) \delta_{qq} \;
\label{eq:Zqq}
\eeq
the ($x$-space) transformations of the NLO and NNLO parton-quark splitting 
functions read
\bea
  ~~\delta [\Delta P_{qq}^{(1)}] &\!\! =\!\! & - \beta_0 z_{qq}^{(1)} 
  \; , \nn \\
  ~~\delta [\Delta P_{qg}^{(1)}] &\!\! =\!\! & 
       z_{qq}^{(1)} \otimes \Delta P_{qg}^{(0)}
\label{eq:dP1} 
\eea
and
\bea
 ~\delta [\Delta P_{qq}^{(2)}] &\!\! =\!\! &
       \beta_0 [( z_{qq}^{(1)})^{\otimes 2} 
       - 2z_{qq}^{(2)}] - \beta_1 z_{qq}^{(1)}  \; , \nn \\
 ~\delta [\Delta P_{qg}^{(2)}] &\!\! =\!\! &
      z_{qq}^{(2)} \otimes \Delta P_{qg}^{(0)}
    + z_{qq}^{(1)} \otimes \Delta P_{qg,L}^{(1)} \: . \;\;
\label{eq:dP2} 
\eea

The non-singlet entries in Eq.~(\ref{eq:Zqq}) can be inferred from the 
fact that the \MSb\ coefficient functions for $g_1^{}$ and the structure 
function $F_3^{\:\!\nu+\bar{\nu}}$ in neutrino-nucleon DIS are identical
to two loops. The functions thus determined agree with the results of the
direct calculation in Ref.~\cite{Matiounine:1998re}. At the moment the
latter article is the only source for the NNLO pure-singlet transformation
term $z_{qq,\rm ps\,}^{(2)}$.

\vspace{1mm}
After transforming to the $x$-space expressions in terms of harmonic 
polylogarithms \cite{Remiddi:1999ew} and performing the transformation 
(\ref{eq:g1trf})$\,$--$\,$(\ref{eq:dP1}), our two-loop expressions fully agree 
with the previous results for $\Delta P_{\rm qf}^{(1)}$ and 
$c_{g_1,\:\!\rm f}^{\:\!(2)}$ discussed in Section~3.
\linebreak \\[-4mm]
Incidentally, we note that the change of $c_{g_1,\,\rm ps}^{\:\!(2)}$ in
the second erratum of \cite{Zijlstra:1993sh} is exactly the scheme 
\linebreak \\[-4mm]  
transformation induced by $z_{qq,\rm ps\,}^{(2)}$ of Ref.~\cite
{Matiounine:1998re}.

\section{Checks and features of the 3-loop results$\!$}

A strong check of the third-order computations is provided by the $\ep^{-3}$ 
and $\ep^{-2\,}$ pole terms of the unfactorized partonic structure functions. 
These terms have to be of a specific form in terms of the lower-order splitting functions and coefficient functions, including the NLO quantities $\Delta 
P_{\!\rm g\,\!f}^{\:\!(1)}$ of Refs.~\cite{Mertig:1995ny,Vogelsang:1995xx} 
inaccessible to a two-loop calculation of photon-exchange DIS. Our results pass
this test.

\vspace{1mm}
We find that $\Delta P_{\rm ns}^{\:\!(2)}$ in Eq.~(\ref{eq:Pexp}) is equal 
to the \linebreak \\[-4mm]
unpolarized non-singlet splitting function $P_{\rm ns}^{\:\!(2)-}$ 
after the scheme transformation $(\ref{eq:dP2})$ adjusting 
$\,c_{g_1^{},\rm ns\,}^{\,(2)}$ to the $(\nu+\bar{\nu})$ coefficient function
$c_{F_3}^{\,(2)\,}$ \cite{F3-NNLO}. 
\\[-0.5mm]
This guarantees that the first moment of $\Delta P_{\rm ns}^{\:\!(2)}$ 
vanishes as required by the conservation of the non-singlet axial vector current.

\vspace{1mm}
Also the first moment of the NNLO gluon-quark splitting function vanishes
\beq
  ~~~\Delta P_{\rm qg}^{\,(2)}(N\! =\! 1) \, = \, 0 
\label{eq:qgN1}
\eeq 
in agreement with the 1990 result of Altarelli and Lampe
\cite{Altarelli:1990jp}. For the pure singlet the relation
\beq
 ~~~\Delta P_{\rm ps}^{\,(n)}(N\! =\! 1) \,=\, 
 - 2\:\!\nf\, \Delta P_{\rm gq}^{\,(n-1)}(N\! =\! 1)
\label{eq:psN1}
\eeq
also holds at NNLO ($n = 2$). Eqs.~(\ref{eq:qgN1}) and (\ref{eq:psN1}) suggest 
the possibility of a higher-order generalisation which would bring us closer to
a complete NNNLO description of $g_1^{}$ at $N=1$.

\vspace{1mm}
Turning to large $x$, we note that the helicity flip is suppressed
by two powers of $(1\! -\! x)$ as $x \ra 1$, 
\beq
 ~~~P_{\rm qg}^{\,(2)}(x) - \Delta P_{\rm qg}^{\,(2)}(x) \:\sim\:
 (1\!-\!x)^2 \cdot P_{\rm qg}^{\,(2)}(x)
\eeq
(the corresponding relation for $\Delta P_{\rm qq}^{\,(2)}$ is obvious as only 
the suppressed pure-singlet parts differ), where $P_{\rm qg}$ denotes the 
unpolarized (flip$\,+\,$non-flip) splitting function. This is in line with the
pattern advocated by Brodsky et al.~\cite{Brodsky:1994kg} for the perturbative and 
non-perturbative contributions to the helicity-dependent parton distributions.
It~is interesting to note that the same suppression holds for 7 of the 8
LO and NLO splitting functions $\,\Delta P_{ij}^{\:\!(0,1)}\!$, the curious
exception being $\Delta P_{\rm gq}^{\:\!(1)}$
(we are not aware of any earlier discussion of this). 
A simple scheme transformation ($z_{gq}^{(1)} \propto \Delta P_{\rm gq}^{(0)})$ 
exists which would `repair' this exception.

\vspace{1mm}
Since it is identical to the unpolarized quantity $P_{\!\rm ns}^{\:\!(2)-}$,
the leading small-$x$ logarithm of $\Delta P_{\rm ns}^{\:\!(2)}$ agrees with
the prediction of the small-$x$ resummation~\cite{SmallxNS}. The situation is
less straightforward in the singlet sector where we find
\bea 
\label{eq:Px0}
  ~~\Delta P_{\rm ps}^{\,(2)} \big|_{\ln^4 x} &\!\! =\!\! & \!
  - \cf \nf\, ( 2\:\!\ca + {\textstyle \frac{8}{3}}\:\! \cf ) \\
  ~~\Delta P_{\rm qg}^{\,(2)} \big|_{\ln^4 x} &\!\! =\!\! & \!
  -5\:\!\cas \nf - {\textstyle \frac{4}{3}}\:\! \cf\nf ( \ca - \nf ) 
  \nn
\eea
($C_A$ and $C_F$ are the usual group factors of QCD).
The first of Eqs.~(\ref{eq:Px0}) agrees with the 1996 prediction of Bl\"umlein
and Vogt \cite{Blumlein:1996hb} based on the singlet resummation by Bartels
et al.~\cite{Bartels:1996wc}. However, the corresponding coefficients of 
$\ca\cf\nf$ and \linebreak \\[-4mm]
$\cfs\nf$ of $\Delta P_{\rm qg}^{\,(2)}$ are not the same. 
Agreement is restored in the supersymmetric limit.

\vspace{1mm}
We have to leave the clarification of this point to further studies. However, 
we should note here that, unlike in the non-singlet case, the leading small-$x$
logs of the singlet splitting functions are not scheme invariant. This can be 
seen from the second lines of Eqs.(\ref{eq:dP1}) and (\ref{eq:dP2}): for 
$z_{qq}^{n} \propto$ \linebreak \\[-4mm]
$c_{g_1^{},\rm q}^{\,(n)}$, for example, $\delta 
[\Delta P_{\rm qg}^{\,(n)}]$ will include leading-log contributions (recall
that $\ln^k x \sim 1/N^{k+1}$ in \mbox{$N$-space}). Hence one cannot conclude 
now that the above difference constitutes an inconsistency.

\vspace{1mm}
A final remark is in order on the $\gamma_5^{}$-induced transformation to the 
\MSb\ factorization scheme. An additional term $z_{gq}^{(2)}$ can be expected 
from the use of, e.g., the Larin scheme in a calculation of an observable 
suitable to access $\Delta P_{\rm gq}^{\,(2)}$ (see below). 
This presently unknown function would add an additional term $ -z_{gq}^{(2)} 
\otimes \Delta P_{\rm qg}^{\,(0)}$ to the first line of Eq.~(\ref{eq:dP2})
without affecting the result for $\Delta P_{\rm qg}^{\,(2)}$.
\pagebreak

Fig.~3 finally illustrates the perturbative expansion of the helicity-%
difference gluon-quark splitting function to the next-to-next-to-leading order.
Here we have employed the code of Ref.~\cite{Gehrmann:2001pz} for the numerical
evaluation of the harmonic polylogarithms \cite{Remiddi:1999ew} up to weight 
four.

\begin{figure}[htbp]
\vspace*{-6mm}
\centerline{\hspace*{-2mm}\epsfig{file=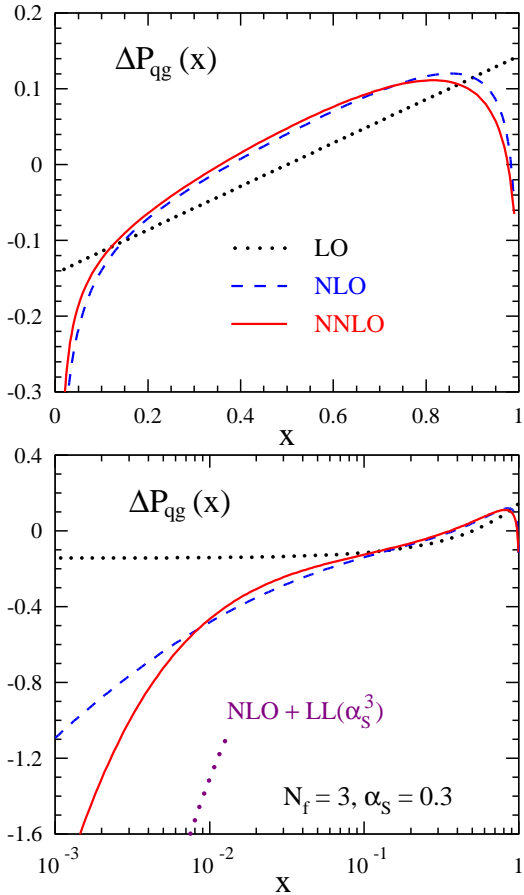,width=7.0cm}}
\vspace{-9mm}
\caption{\label{Pqg-NNLO}
The LO, NLO, and NNLO approximations for the splitting function 
$\Delta P_{\rm qg}$ at values of $\as$ and the number $\nf$ of light flavours
relevant to present data on the spin structure function $g_1^{}$.}
\vspace*{-6mm}
\end{figure}

The third-order corrections are 15\% or less over the wide region 
$0.005 \leq x \leq 0.9$ even for this relatively large value of $\as$.
The figure also shows that, as for all other three-loop quantities we have
calculated \cite{MVV-Pij,MVV-proc,MVV-F2L} (see also Refs.~\cite{Moch:2007tx}
for the time-like case), 
the leading small-$x$ logarithm
does not provide a useful approximation to the complete function at any 
relevant values of~$x$.

\section{Summary and Outlook}

We have calculated the three-loop QCD corrections to spin-dependent 
electromagnetic deep-inelastic scattering and extracted the upper row 
$\Delta P_{\rm qf}^{\,(2)}(x)$, $f = q,g$, of the matrix of the third-order 
(NNLO) spin splitting functions.

\vspace{0.5mm}
Due to the presence of difficult tensor integrals beyond those encountered
before \mbox{\cite{MVV-Pij,MVV-proc,MVV-F2L}}, the computation of the 
corresponding diagrams provided yet another example of the `no free 
lunch' theorem. 
The $D \neq 4$  continuation of $\gamma_5^{}$ has been dealt with using the
Larin prescription \cite{g5-Larin} and, as in previous two-loop calculations 
\cite{Mertig:1995ny,Vogelsang:1995xx}, a final scheme transformation to 
\MSb\ factorization. The result for $\Delta P_{\rm qg}^{\,(2)}$ is final also 
in this respect, while the pure-singlet quark splitting function
\linebreak
$\Delta P_{\rm ps}^{\,(2)}$ may require an additional transformation.

\vspace{0.8mm}
Our calculation also includes the coefficient functions for the spin structure 
function $g_1^{}$ up to the NNNLO (again in \MSb\ up to a scheme 
transformation in the pure-singlet sector). At~two loops we confirm the
previous results for the NNLO coefficient functions 
$c_{g_{1\!}^{},\rm f}^{(2)}$ of Ref.~\cite{Zijlstra:1993sh} which were not 
independently checked before.
Returning to the NNLO splitting functions, we note that our results exhibit
interesting features at the moment $N=1$ and the momentum-fraction limits 
$x \ra 1$ and $x \ra 0$ which merit further studies.

\vspace{0.8mm}
We have started work on the lower row $\Delta P_{\rm gf}^{\,(2)}$, $f = q,g$
of the splitting-function matrix. Our method is a generalisation of the 
scalar-exchange approach used in the unpolarized case~\cite{MVV-Pij,MVV-proc}, 
i.e., we now consider DIS including also the exchange of a pseudoscalar 
(Higgs in the heavy-top limit)~$\chi$ coupling to 
${\widetilde G}^a_{\!\mu\nu} G_{\!a}^{\mu\nu}$. 
This adds an operator mixing to the problem~\cite{Chetyrkin:1998mw}. Moreover,
there are many more diagrams in this sector, e.g., a preliminary 
(non-optimised) database has about 30000 $g \phi\:\! g \chi\,$ diagrams where
$\phi$ is the scalar Higgs used before. 

\vspace{1mm}
Despite the expected impact of improved tools such a the new multi-threaded 
version of {\sc Form} \cite{Tentyukov:2007mu}, it will thus take some time
before all NNLO spin splitting functions are known. In the meantime, the full 
$N$- and $x$-space results for $\Delta P_{\rm qf}^{\,(2)}$ (including
{\sc Form} and {\sc Fortran} routines) will be presented in a forthcoming 
journal publication.

\section*{Acknowledgements}

We would like to thank W.~Vogelsang for useful discussions.
The work of A.V. has been supported in part by the UK Science \& Technology
Facilities Council under grant number PP/E007414/1.
S.M. and M.R. acknowledge support by the Helm\-holtz Gemeinschaft under 
contract VH-NG-105 and by the Deutsche Forschungsgemeinschaft in 
Sonderforschungs\-be\-reich/Transregio~9.
The work of J.V. has been part of the research program of the Dutch
Foundation for Fundamental Research of Matter (FOM).

\end{document}